\documentclass[twocolumn,showpacs,prb,floatfix,superscriptaddress,citeautoscript
,cite]{revtex4}
\usepackage{graphicx}
\usepackage{epstopdf}
\usepackage{color}
\usepackage{amsmath}
\usepackage{ulem}

\begin{document}

\title{Distinctive magnetic properties of CrI$_3$ and CrBr$_3$ monolayers \\caused by spin-orbit coupling}

\author{C. Bacaksiz}
\affiliation{Department of Physics, University of Antwerp, Groenenborgerlaan 171, B-2020 Antwerp, Belgium}
\affiliation{NANOlab Center of Excellence, University of Antwerp, Belgium}
\affiliation{Bremen Center for Computational Material Science (BCCMS), Bremen D-28359, Germany}
\affiliation{Computational Science Research Center, Beijing and Computational Science and Applied Research Institute Shenzhen, Shenzhen, China}

\author{D. \v{S}abani}
\affiliation{Department of Physics, University of Antwerp, Groenenborgerlaan 171, B-2020 Antwerp, Belgium}
\affiliation{NANOlab Center of Excellence, University of Antwerp, Belgium}

\author{R. M. Menezes}
\affiliation{Department of Physics, University of Antwerp, Groenenborgerlaan 171, B-2020 Antwerp, Belgium}
\affiliation{NANOlab Center of Excellence, University of Antwerp, Belgium}
\affiliation{Departamento de Física, Universidade Federal de Pernambuco, Cidade Universitária, 50670-901, Recife-PE, Brazil}

\author{M. V. Milo\v{s}evi\'c}
\email{milorad.milosevic@uantwerpen.be}
\affiliation{Department of Physics, University of Antwerp, Groenenborgerlaan 171, B-2020 Antwerp, Belgium}
\affiliation{NANOlab Center of Excellence, University of Antwerp, Belgium}

\begin{abstract}
After the discovery of magnetism in monolayer CrI$_3$, the magnetic properties of different 2D materials from the chromium-trihalide family are intuitively assumed to be similar, yielding magnetic anisotropy from the spin-orbit coupling on halide ligands. Here we reveal significant differences between the CrI$_3$ and CrBr$_3$ magnetic monolayers in their magnetic anisotropy, resulting Curie temperature, hysteresis in external magnetic field, and evolution of magnetism with strain, all predominantly attributed to distinctly different interplay of atomic contributions to spin-orbit coupling in two materials. 
\end{abstract}

\date{\today}
\pacs{Valid PACS appear here}
\maketitle

\section{Introduction} 

In the family of two-dimensional (2D) materials, exhibiting a range of exciting and advanced properties\cite{Park_2016}, the intrinsic magnetism was long evasive. The first pathway to realize the magnetic 2D crystal has been exfoliation from layered bulk magnets. Nearly a decade after realization of graphene, exfoliated monolayer FePSe$_3$\cite{lee2016ising,wang2016raman} and CrSiTe$_3$\cite{lin2016ultrathin} were reported to have long-range magnetic order, indirectly demonstrated via Raman and conductivity measurements, respectively. However, the field of 2D magnetism has truly boomed only after the premiere direct evidence of two-dimensional (2D) ferromagnetism, in monolayer CrI$_3$\cite{huang2017layer} and bilayer Cr$_2$Ge$_2$Te$_6$\cite{gong2017discovery}, that attracted much attention of the scientific community. Since then, a number of new 2D magnets were synthesized and utilized in different heterostructures\cite{gibertini2019magnetic}, including monolayer CrBr$_3$\cite{kim2019micromagnetometry} and CrCl$_3$\cite{klein2019enhancement}, other members of the Cr-trihalide family next to CrI$_3$.

Although the number of 2D magnets exfoliated either from magnetic or nonmagnetic\cite{bonilla2018strong,fei2018two,o2018room} bulk counterparts is increasingly large, the physics behind the magnetism in these materials is common. Namely, the magnetic moment originates from unpaired $d$-electron of the transition metal atoms, but the stability of magnetization at finite temperature comes as the consequence of magnetic anisotropy, lifting the restrictions stipulated by Mermin-Wagner theorem\cite{Mermin66}. There are two possible sources of the magnetic anisotropy in 2D materials: one is the anisotropy of the magnetic ion due to the character and symmetry of bond coordination with the non-magnetic atoms, known as single-ion anisotropy (SIA); the other is the anisotropy in magnetic exchange interaction between the magnetic atoms within the crystal. In each case, the spin-orbit coupling (SOC) is a direct responsible for the arising magnetic anisotropy\cite{PhysRev.120.91}. 

Before the experimental realization of monolayer CrI$_3$, the chromium-halides were predicted to be ferromagnetic in the monolayer form\cite{liu2016exfoliating}, where by including SOC in the consideration, magnetic anisotropy energies (MAEs) were calculated\cite{zhang2015robust}. After the actual exfoliation of CrI$_3$, a more detailed study on the chromium-halides reported the tunability of MAE by strain\cite{PhysRevB.98.144411}. Other studies discussed accurate calculation of the critical temperature of CrI$_3$ and other 2D magnetic materials using spin-wave theory and Monte-Carlo (MC) simulations on top of \textit{ab initio} results\cite{torelli2019high,torelli2018calculating,olsen2019theory}, and were mostly focused on the proper description of the magnetic interactions. Very recent works then considered adsorption and substitution of foreign atoms, such as hydrogen or oxygen\cite{rassekh2020remarkably,pizzochero2020atomic,pizzochero2020inducing}, to manipulate the exchange interaction in order to increase the critical temperature. 

\begin{figure*}[t]
\includegraphics[width=0.7\linewidth]{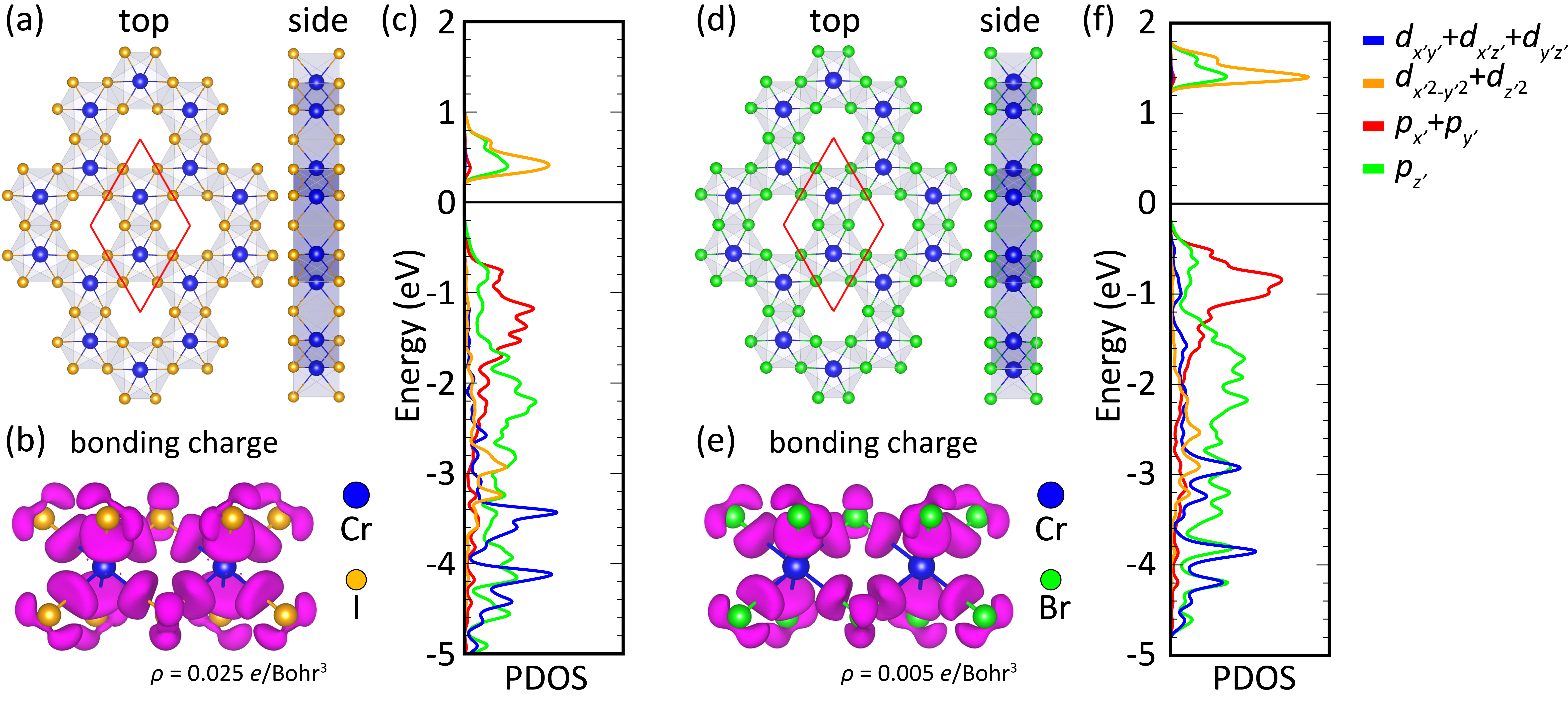}
\caption{\label{f1} Schematic representation of the structure of monolayer CrI$_3$ (a) and CrBr$_3$ (d). Panels (b) and (e) show the difference between the charge distribution after crystallization and the total charge distribution of bare atoms, which then indicates the bonding and anti-bonding charges in the two materials. (c) and (f) panels show the density of states of two materials, decomposed according to the atomic orbitals. Subscripts $x'$, $y'$ and $z'$ in the orbitals denote the local coordinates of the corresponding atoms.} 
\end{figure*}

It is now well established that the origin of the magnetic anisotropy in monolayer CrI$_3$, for both SIA and anisotropy of the exchange interactions, is associated with the SOC of iodine atoms rather than chromium ones\cite{lado2017origin,xu2018interplay,PhysRevB.101.134418,PhysRevLett.122.207201}. As a consequence one can intuitively predict that the magnetic anisotropy of monolayer CrBr$_3$ should be lower as compared to that in CrI$_3$ due to difference in SOC between I and Br. As a corroboration to these predictions, recent experiment established the Curie temperature of CrBr$_3$ as 21 K\cite{kim2019micromagnetometry}, significantly lower than one of monolayer CrI$_3$ ($T_C=45$ K\cite{huang2017layer}). However, the extent and manner of how individual atomic contributions to SOC affect the magnetic anisotropy in different monolayer magnets remained unaddressed to date. Therefore, in this article we perform a thorough comparison between seemingly similar monolayer magnets, CrI$_3$ and CrBr$_3$, in order to provide a comprehensive (qualitative and quantitative) understanding of the atomically-resolved effects of spin-orbit coupling on magnetism. Specifically, we explore the magnetic anisotropy, variation of magnetism under biaxial strain, critical temperature and response to the magnetic field, and link the differences in SOC in two similar materials to their rather dissimilar emergent magnetic properties.

The paper is organized as follows. In Sec. \ref{comp} we provide details of our computational methodology. The structural and electronic properties, the magnetic properties, and the effect of strain are presented and discussed in Sec. \ref{struc}, \ref{magn}, and \ref{strain}, respectively. In Sec. \ref{temp-hys}, the temperature-dependent magnetism and magnetization reversal of the two materials in external magnetic field are discussed. Sec. \ref{conc} summarizes our findings.

\section{Computational Methodology}\label{comp}

In order to investigate the structural, electronic and magnetic properties we use calculations based on density functional theory (DFT). To obtain magnetic parameters, we employed four-state energy mapping methodology\cite{xiang2013magnetic,sabani2020ab}. When different magnetic configurations are examined, the magnetic moments are constrained in desired directions, in order to prevent their relaxation into the ground state configuration (or any stable configuration other than desired one) during the self-consistent procedure. Heisenberg spin Hamiltonian is considered in the form:
\begin{equation}\label{hamil}
 H=\frac{1}{2}\sum_{i , j} \mathbf{S}_i \mathbf{J}_{ij} \mathbf{S}_j + \sum_{i} \mathbf{S}_i \mathbf{A}_{ii} \mathbf{S}_i,
\end{equation}
where $\mathbf{S}_i = (S_{i}^{x},S_{i}^{y},S_{i}^{z})$ is a vector. $\mathbf{J}_{ij}$ and $\mathbf{A}_{ii}$ are $3 \times 3$ matrices describing the magnetic exchange interaction between the different sites and the single-ion anisotropy (SIA), respectively.

We used the Vienna \textit{ab initio} simulation package VASP\cite{vasp1,vasp2,vasp3} which solves the Kohn-Sham equations iteratively using a plane-wave basis set. To describe electron exchange and correlation, the Perdew-Burke-Ernzerhof (PBE) form of the generalized gradient approximation (GGA)\cite{pbe} was adopted. The spin-orbit coupling (SOC) was included in all calculations, as VASP includes SOC via following added term\cite{PhysRevB.93.224425} to DFT Hamiltonian:  
\begin{equation}\label{soc-ham}
 H_{SO}= \frac{\hbar^2}{(2 m_{e} c^{2})^{2}} \bigg( 1- \frac{V(r)}{2 m_{e} c^{2} } \bigg)^{-2} \frac{dV(r)}{dr} \vec{\sigma} \cdot \vec{L},
\end{equation}
where $\vec{\sigma}= (\sigma_{x},\sigma_{y},\sigma_{z})$ stands for $2\times2$ Pauli spin matrices, $\vec{L} = \vec{r}\times\vec{p}$ is the angular momentum, and $V(r)$ is the spherical part of the electron potential. The van der Waals (vdW) forces were taken into account using the DFT-D2 method of Grimme\cite{grimme_d2}. In order to calculate charge transfer between the atoms we employed the Bader charge technique\cite{henkelman2006fast}.

The kinetic energy cut-off of the plane-wave basis set was 600 eV and energy convergence criterion was 10$^{-6}$ eV in the ground-state calculations. Gaussian smearing of 0.01 eV was used and the pressures on the unit cell were decreased to a value lower than 1.0 kbar in all three directions. On-site Coulomb repulsion\cite{PhysRevB.57.1505} parameter, $U$, was taken as 4 eV for magnetic Cr atom.\cite{sivadas2015magnetic,chittari2016electronic,PhysRevB.99.144401} To avoid interactions between periodically repeating monolayers in vertical direction, our calculations were performed with sufficiently large vacuum space of $\sim$13 \AA{}. 

For subsequent considerations of the temperature-dependent magnetization and the critical temperature ($T_\text{C}$), we performed spin dynamics simulations based on stochastic Landau-Lifshitz-Gilbert (LLG) equation, using simulation package $Spirit$\cite{PhysRevB.99.224414}, adapted to accommodate the anisotropic interactions of our Hamiltonian [Eq.~\eqref{hamil}]. 50$\times$50 supercell of the spin lattice was considered. The spin system is initialized in random configuration at high temperature and then cooled down with the temperature steps of 0.125 K. To obtain equilibrium magnetization took relaxation over $10^5$ time steps (with time step $\Delta t=1$ fs) for each temperature. For the field-dependent calculations, the Zeeman term $H_B=g\mu_B\sum_i\textbf{S}_i\cdot\textbf{B}$ has been included in Eq.~\eqref{hamil}, where $\textbf{B}$ is the applied magnetic field; $g=2$ is the g-factor for $S=3/2$; $\mu_B$ the Bohr magneton and $\mu=gS\mu_B=3\mu_B$ is the magnetic moment of Cr atoms.

\begin{table}[htbp]
\caption{\label{struc_tab} Structural and electronic parameters of monolayer CrI$_3$ and CrBr$_3$. The charge transfer per atom ($\rho_{Cr}$ and $\rho_{X}$) was calculated using Bader charge technique.}
\begin{tabular}{lcccccccc}
\hline\hline
           & $d_b$  & $a$    & $\theta$     & $\rho_{Cr}/\rho_{X}$  & $E_g$  \\
           & (\AA{})& (\AA{})  & ($^{\circ}$) & ($e^-$ / $e^-$)  & (eV)     \\
\hline
CrI$_3$    & 2.78   & 6.94     & 92.06         & 1.10/$-0.37$ & 0.55     \\
CrBr$_3$   & 2.55   & 6.40     & 92.90         & 1.33/$-0.44$ & 1.59     \\
\hline\hline 
\end{tabular}
\end{table}

\section{Results and Discussion}\label{r-and-d}

\subsection{Structural and electronic properties}\label{struc}

We start from the structural properties of monolayer CrX$_3$, where X stands for ligand I and Br atoms. CrX$_3$ crystallizes in the trigonal $P\overline{3}1m$ space group. The hexagonal planar lattice of Cr atoms is sandwiched between triangular planar lattices of ligand atoms as shown in Figs. \ref{f1} (a) and (d) for respective materials. One Cr atom bonds to 6 ligand atoms and each ligand atom bonds to 2 Cr atoms. The structures have 3-fold in-plane symmetry, where the triangular lattices of ligand atoms are 180$^{\circ}$ rotated with respect to each other, which creates octahedral coordination around each Cr atom. As listed in Table \ref{struc_tab}, the Cr-I and Cr-Br bond lengths are found to be 2.78 \AA{} and 2.55 \AA{}, respectively, which is directly proportional with the radius of the bonding orbital of the ligand, I-5$p$ and Br-4$p$. Consequently, the lattice constants of monolayer CrI$_3$ and CrBr$_3$ are also different, 6.94 \AA{} and 6.40 \AA{}, respectively, consistent with the experimentally measured 6.95 \AA{}\cite{li2020single} and 6.50 \AA{}.\cite{chen2019direct} Note that both monolayers exhibit slightly larger lattice constant than their bulk or few-layer counterparts. The covalent bonding character consists of 1.10 $e^-$ and 1.33 $e^-$ donation of Cr and 0.37 $e^-$ and 0.44 $e^-$ gain of I and Br atoms, respectively. This bonding charge is visualized in Figs. \ref{f1} (b) and (e), as obtained by subtracting bare atom charge distributions from the charge distribution of the crystals. Charges obviously accumulate to the bonding sites and around Cr atoms. Observed charge accumulation around each ligand clearly shows three 'charge  clouds’ that can be divided in two groups,  representing two types of bonding between Cr and ligand atoms: the facial ($\sigma$) bonding, represented by two charge clouds facing the Cr atoms, and the lateral ($\pi$) bonding, represented by the third charge cloud, in direction orthogonal to the Cr-X-Cr-X plane.

To depict the energetic arrangements of the orbital states, we calculated the partial density of states for two considered materials in Figs. \ref{f1} (c) and (f). Both materials are semiconductors, in which the valence band maximum is dominated by $p$-orbital of the ligand. On the other hand, $d$-orbital of Cr resides much deeper in the valence band, with similar energetic delocalization for both CrI$_3$ and CrBr$_3$. $d_{x'y'}$, $d_{x'z'}$ and $d_{y'z'}$ are degenerate, and are plotted together. $d_{z'^2}$ and $d_{x'^2 - y'^2}$ orbitals are also degenerate and appear at the conduction band. $p_{x'}$ and $p_{y'}$ orbitals are degenerate as well, and dominate the valence band maximum. $p_{z'}$ orbital mostly resides in the middle of the valence band. One should note that in order to obtain orbital states compatible with the octahedral coordination, the general coordinates are rotated as suggested by Rassekh \textit{et al.}\cite{rassekh2020remarkably}. Therefore we consider $xy$-plane and $z$-axis of the general Cartesian coordinates aligned with the local coordinates, where the Cr-X-Cr-X plane is considered as $x'y'$-plane, and $z'$ axis is orthogonal to that plane. Orbital decomposition shows that both materials exhibit very similar orbital delocalizations. This is also visible in the charge density variations shown in Figs. \ref{f1} (b) and (e). Purple regions show the depletion of charges of isolated atoms through the charge density of the crystal. It is obvious that the orbitals laying along the Cr-X bonds exhibit most delocalization. We understand from the charge depletion and the orbital decomposition of DOS that $d_{x'y'}$, $d_{x'z'}$ and $d_{y'z'}$ orbitals are localized and do not show any variation from their single-atom form. $d_{z'^2}$ and $d_{x'^2 - y'^2}$ orbitals form a $spd^2$ hybridization with the $p_{x'}$ and $p_{y'}$ orbitals as expected for an octahedral coordination. Beside these similarities, the particular difference between CrI$_3$ and CrBr$_3$ is found between 4$p$ of Br and 5$p$ of I, leading to energetic differences of the $spd^2$ hybridization. Since 4$p$ orbital of Br is more confined as compared to 5$p$ of I, the states of CrBr$_3$ are shifted to higher energies. Therefore, the band gap values are different, found as 0.55 eV and 1.59 eV for CrI$_3$ and CrBr$_3$, respectively.

\begin{figure}[t]
\includegraphics[width=\linewidth]{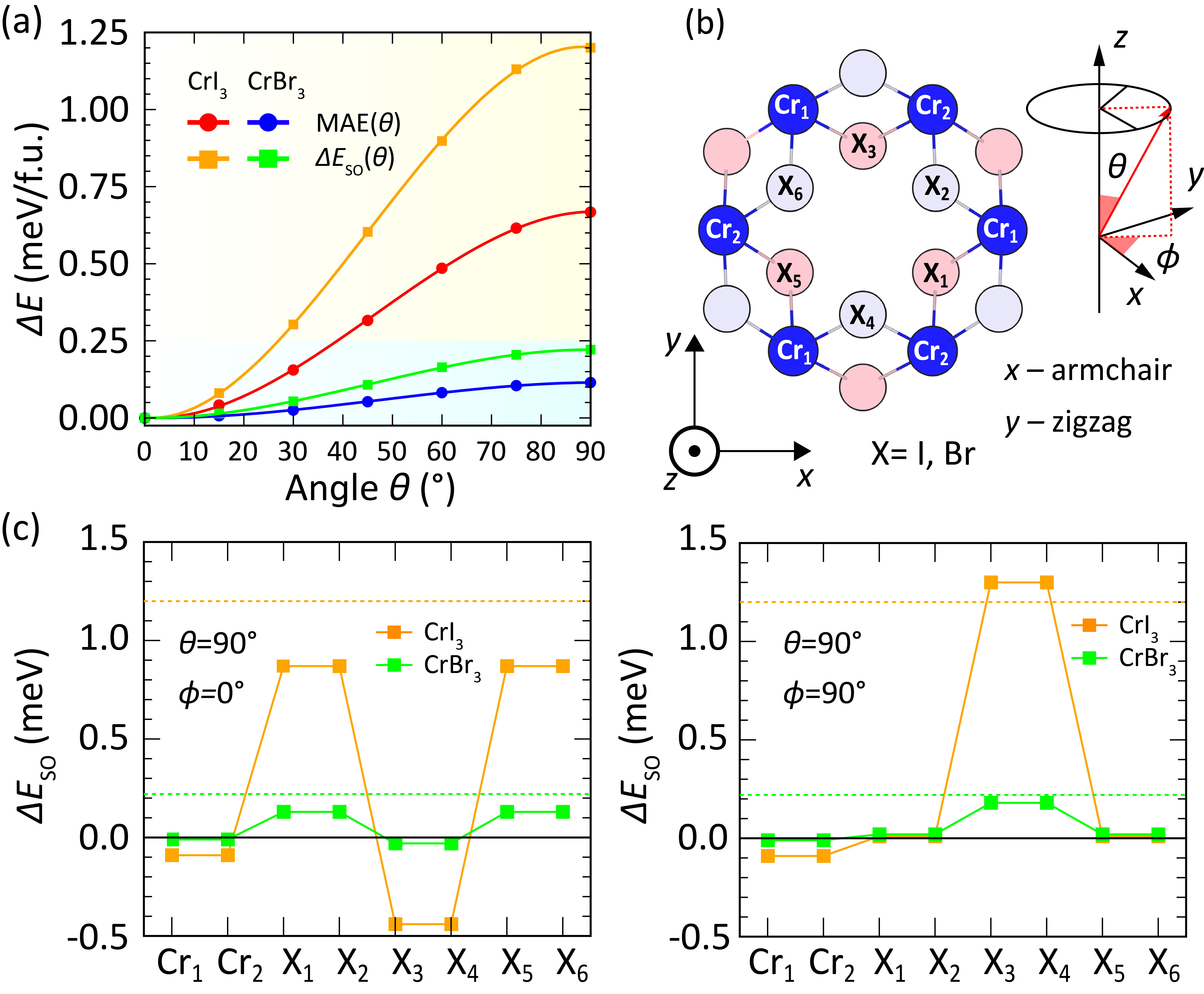}
\caption{\label{f2} (a) Magnetic anisotropy energy (MAE) and SOC energy as a function of the spin alignment angle with the out-of-plane direction. (b) Schematic representation of the top view of the structure. Atoms are enumerated in order to track the variation of the corresponding SOC energies for spin-alignment in $x$-direction ($\theta = 90^\circ$ and $\phi = 0^\circ$) (c), and in $y$-direction ($\theta = 90^\circ$ and $\phi = 90^\circ$) (d). The energy of the out-of-plane spin alignment, the ground state, is set to 0 eV. Dashed lines represent total $\Delta E_{\text{SO}}$, while square dots (connected by solid lines) represent contributions of each atom to the total $\Delta E_{\text{SO}}$.} 
\end{figure}

\subsection{Magnetic properties}\label{magn}

\subsubsection{Spin-orbit coupling and magnetic anisotropy energy}

It is already known that monolayer CrI$_3$ exhibits ferromagnetism at finite temperature due to the magnetic anisotropy originating from SOC on I atoms. It is natural to assume that a similar mechanism is responsible for ferromagnetism in CrBr$_3$, as proposed in previous studies\cite{PhysRevB.98.144411}. In what follows we validate that assumption, but more importantly, we analyze deeper the differences in individual atomic contribution to the total SOC in monolayer CrI$_{3}$ and CrBr$_{3}$, and the consequences thereof.

First of all, we calculate the total energy leading to magnetic anisotropy energy (MAE) and the contributed SOC energy ($\Delta E_{\text{SO}}$), relative to the respective ground-state energies, depending on the spin direction angle ($\theta$) with respect to the out-of plane direction. 
In other words, $\text{MAE}(\theta) = E(\theta)-E(\theta = 0^\circ)$ and $\Delta E_{\text{SO}}(\theta) = E_{\text{SO}}(\theta)-E_{\text{SO}}(\theta = 0^\circ)$, where angle $\theta$ is measured from $z$-axis (see Fig. \ref{f2}(b)). MAE is found to be 0.67 (0.11) meV for for CrI$_3$ (CrBr$_3$) favoring the out-of-plane direction. In the literature, the reported MAE values vary from $\sim 0.69 (0.18)$ meV to $\sim1.48 (0.30)$ meV for CrI$_3$ (CrBr$_3$) depending on the approximations used.\cite{zhang2015robust,PhysRevB.98.144411,Albaridy} However, MAE of CrI$_3$ is 4-5 times larger than that of CrBr$_3$ in each previous report, as is in the present study. 
As shown in Fig. \ref{f2}(a), MAE$(\theta)$ and $\Delta E_{\text{SO}}(\theta)$ exhibit sinusoidal functional behavior. It is further remarkable that MAE$(\theta)$ is not equal to $\Delta E_{\text{SO}}(\theta)$, which indicates that the collective rotation of the spins allows further relaxation of the spatial part of the wave function, to a lower energy. The ratio between MAE$(\theta)$ and $\Delta E_{\text{SO}}(\theta)$ is around 0.5 for either monolayer, at each angle $\theta$. 

Furthermore, we calculate the atomic contributions to the total SOC energies for the spins pointing in $x$- ($\theta=90^\circ$ and $\phi=0^\circ$) and $y$-direction ($\theta=90^\circ$ and $\phi=90^\circ$), as shown in Figs. \ref{f2} (c) and (d). The total contribution from Cr atoms is significantly lower than that from ligand atoms, which confirms results of previous works\cite{lado2017origin,xu2018interplay}. More importantly, the atomic contributions of both I and Br vary depending on the direction of the bond coordination of the ligand with respect to the spin direction. To be more direct, in Fig. \ref{f2}(c) the ligands X$_1$, X$_2$, X$_5$, and X$_6$ energetically prefer the spin pointing in the $z$-direction, while X$_3$ and X$_4$ prefer the spin in $x$-direction. On the other hand, in Fig. \ref{f2}(d), the contributions from X$_1$, X$_2$, X$_5$, and X$_6$ are almost zero, however, X$_3$ and X$_4$ prefer the spins aligned with $z$-direction. Briefly, the contribution $\vec{\sigma} \cdot \vec{L}$ depends on the direction of the orbital angular momentum of the ligand $L_{x'}$, $L_{y'}$, which lay on the respective Cr-X bonding directions as $p_{x'}$ and $p_{y'}$. There is interplay with the $d_{z'^2}$ and $d_{x'^2 - y'^2}$ orbitals through the $spd^2$ hybridization. On the other hand, $L_{z'}$ is orthogonal to Cr-X-Cr-X plane. These results not only confirm the previous works suggesting the main contribution of MAE originates from SOC of the ligand atoms, but also reveal that the main contributions are stemming from the bonding orbitals. Previously, Lado \textit{et al.} reported MAE of monolayer CrI$_3$ as a function of SOC strength on Cr and I (ligand) separately and concluded that the SOC on ligand is the dominant factor on the MAE.\cite{lado2017origin} Here we further reveal that not every ligand has positive contribution to MAE - those with a bond with Cr atom aligned with magnetization direction of interest, will have negative (or no) contribution.  The apparent difference between CrI$_3$ and CrBr$_3$, on the other hand, is a direct consequence of the relative strength of the SOC of 5$p$ orbital of the I atom compared to the 4$p$ orbital of the Br atom.

\begin{table*}[t]
\caption{\label{mag_tab} Magnetic exchange parameters of monolayer CrI$_3$ and CrBr$_3$. $J^{xx}$, $J^{yy}$, and  $J^{zz}$ are diagonal elements, and $J^{xy} = J^{yx}$, $J^{xz} = J^{zx}$, $J^{yz} = J^{zy}$ are off-diagonal elements of the exchange matrix. The mean value $\langle J \rangle$ of the exchange parameters is also given. Out-of-plane anisotropy $\Delta$ is calculated as $\langle J^{xx}\rangle - \langle J^{zz}\rangle$. $A_{ii}$ is SIA parameter, same for each Cr site. MAE and $E_{\text{SO}}$ are magnetic anisotropy and total SOC energies, respectively.}
\begin{tabular}{lcccccccccccccc}
\hline\hline
  &  pair     &$J^{xx}$& $J^{yy}$ & $J^{zz}$ & $J^{xy} = J^{yx}$ & $J^{xz} = J^{zx}$  & $J^{yz} = J^{zy}$&$\Delta$& $A_{ii}^{zz}$& MAE & $\Delta E_{\text{SO}}$ \\
  &  ($i$-$j$)& (meV) & (meV)   & (meV)  & (meV)              &    (meV)         &    (meV) & (meV) &(meV)           & (meV/f.u.)&   (meV/f.u.) \\
\hline
CrI$_3$&                   &       &         &        &                     &                   &      & -0.22 &  -0.07  & 0.67  &  1.21  \\
       & (1-2)  & -5.10 &  -3.72  &  -4.63 &   0.00              & 0.00              & 0.83 & &        &       &          \\
       & (2-3)  & -4.07 &  -4.76  &  -4.63 &  -0.60              & 0.72              &-0.42 & &        &       &          \\
       & (2-5)  & -4.07 &  -4.76  &  -4.63 &   0.60              &-0.72              &-0.42 & &        &       &          \\
&$\langle J \rangle$     & -4.41 &  -4.41  &  -4.63 &   0.00              & 0.00              & 0.00 &       &         &       &          \\
CrBr$_3$&                   &       &         &        &                     &                   &      & -0.04 & -0.01  & 0.11  & 0.22   \\
       & (1-2) & -3.45 &  -3.29  &  -3.42 &   0.00              & 0.00              & 0.09 & &        &       &        \\
       & (2-3)  & -3.33 &  -3.41  &  -3.42 &  -0.07              & 0.08              &-0.05 & &        &       &         \\
       & (2-5)  & -3.33 &  -3.41  &  -3.42 &   0.07              &-0.08              &-0.05 & &        &       &               \\
&$\langle J \rangle$& -3.37 &  -3.37  &  -3.42 &   0.00              & 0.00              & 0.00 &       &        &       &        \\
\hline\hline 
\end{tabular}
\end{table*}

\subsubsection{Magnetic exchange interaction and SIA}

\begin{figure}[b]
\includegraphics[width=\linewidth]{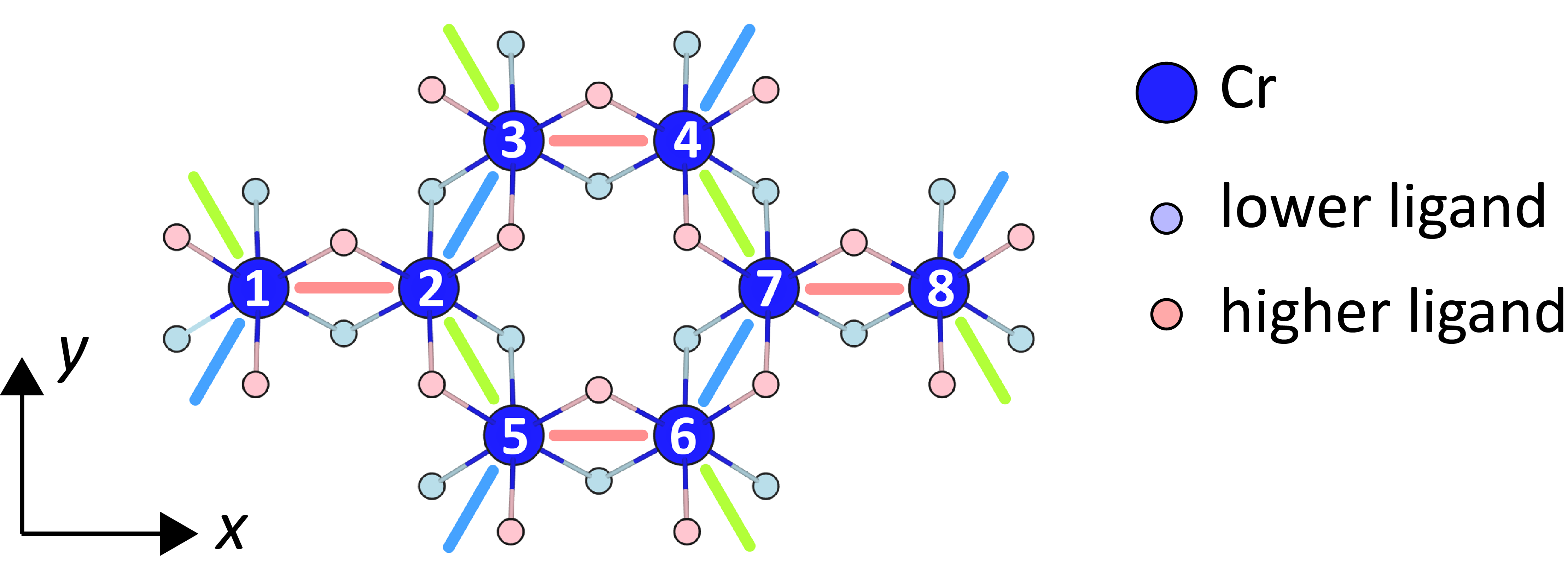}
\caption{\label{fx} $2\times2$ supercell where Cr atoms are indexed consistently with Table \ref{mag_tab}. The equivalent Cr-Cr bonds are shown with the same color.} 
\end{figure}

To characterize the magnetic properties of the two monolayers under study, we calculate the magnetic exchange interactions matrix ($\mathbf{J}_{12}$) in the first nearest-neighbor (NN) approximation, using a $2 \times 2$ supercell shown in Fig. \ref{fx}. Due to the threefold in-plane symmetry of CrX$_3$, once the exchange matrix of one of the three first NN pairs is obtained, such as for the pair (1-2), the matrices of other first NN pairs, (2-3) and (2-5), can be calculated by rotation operation on the exchange matrix of pair (1-2) around the out-of-plane axis. The so obtained pairwise results are listed in Table \ref{mag_tab}. Depending on the considered coordinate axes, the exchange matrix of a pair can change. Here (1-2) pair lays on the $x$-axis, consequently (2-3) and (2-5) pairs make $60^\circ$ and $-60^\circ$ angle with the $x$-axis. Each pair of CrI$_3$ and CrBr$_3$ has a symmetric exchange matrix, due to preserved inversion symmetry, hence no Dzyaloshinskii-Moriya interaction (DMI) is present\cite{sabani2020ab}. All diagonal elements show that both monolayer materials exhibit ferromagnetic interaction in the first NN consideration. As an effective exchange interaction for Cr, the mean of the matrices of (1-2), (2-3), and (2-5) pairs is listed in Table \ref{mag_tab} as $\langle J \rangle$, and is identical for each magnetic site in the respective monolayer. In this form of representation, it is clearly seen that CrI$_3$ exhibits stronger ferromagnetic exchange as compared to CrBr$_3$. It is also rather remarkable that the out-of-plane exchange anisotropy, $\Delta= \langle J \rangle^{xx}-\langle J \rangle^{zz}= \langle J \rangle^{yy}-\langle J \rangle^{zz}$ of CrI$_3$  is 0.22 meV, much larger than 0.04 meV of CrBr$_3$. Finally, we also calculated SIA, which as a result of threefold symmetry is represented by a single parameter, $A_{ii}^{zz}$, out of nine elements of the $\mathbf{A}_{ii}$ matrix. $A_{ii}^{zz}$ of CrI$_3$ is found to be $-0.07$ meV, much larger than that of CrBr$_3$, $-0.01$ meV. Both monolayers exhibit negative SIA, which indicates the energetic preference of the spin alignment in the out-of-plane direction. It is worth to mention that Xu \textit{et al.} previously calculated $3 \times 3$ exchange interaction using different set of calculation parameters for CrI$_3$ and therefore reported different values,\cite{xu2018interplay} however, our results are qualitatively consistent with their ones. Namely, they found the ferromagnetic diagonal elements with anisotropy favoring the out-of-plane direction, symmetric off-diagonal elements, and SIA parameter of Cr atoms favoring out-of-plane as well. They also analysed the exchange and SIA parameters as a function of SOC strength and concluded the main contribution to those parameters comes from the SOC in I atoms as Lado \textit{et al.}\cite{lado2017origin} did.

In order to reveal the effect of SOC from another perspective, we next calculated the exchange matrix and SIA parameter in the non-collinear scheme without SOC. As expected, the exchange matrix for both monolayers is diagonal, with equal diagonal elements. SIA parameter is zero, which is also expected. The results confirm that the origin of the magnetic anisotropy, consequently the origin of the magnetization of these monolayer materials, is the spin-orbit interaction. However the comparison between exchange parameters with and without SOC further indicates the difference between CrI$_3$ and CrBr$_3$. The exchange parameter of CrI$_3$ without SOC is found to be $-4.53$ meV which is between $\langle J \rangle^{xx}=\langle J \rangle^{yy}=-4.41$ meV and $\langle J \rangle^{zz}=-4.63$ meV when SOC is included. For CrBr$_3$, on the other hand, it is found to be $-3.38$ meV which is almost equal to the in-plane exchange $\langle J \rangle^{xx}=\langle J \rangle^{yy}=-3.37$ meV. Therefore, one understands that the SOC in CrI$_3$ not only enhances the exchange interaction of the out-of-plane spin components but also reduces the interaction of the in-plane components. For CrBr$_3$, the SOC contributes to the interaction of out-of-plane spins only.

\begin{figure*}[t]
\includegraphics[width=0.85\linewidth]{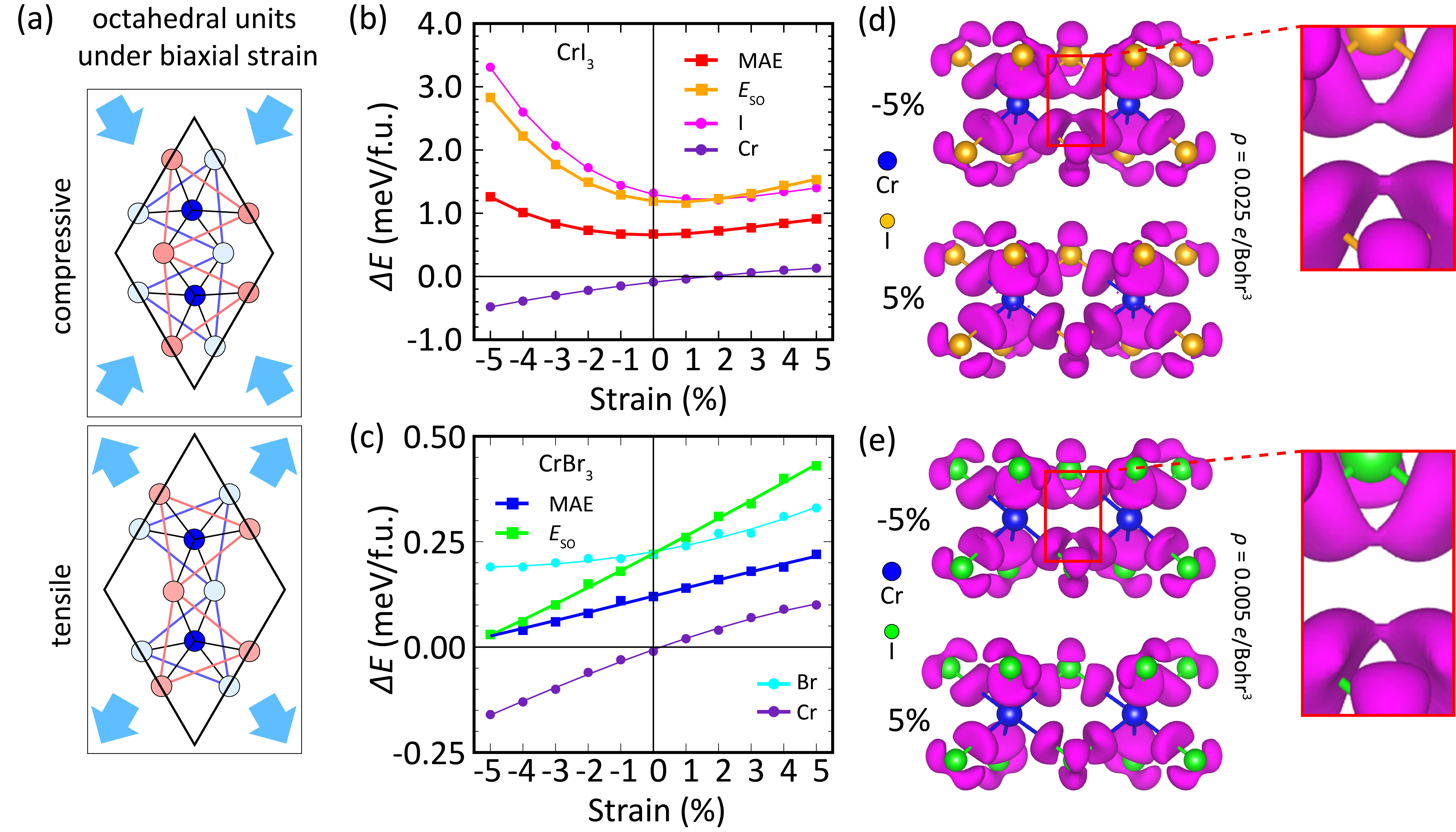}
\caption{\label{f3} (a) Schematic illustration of the distortion caused on the octahedral unit of the material under tensile and compressive strain. Panels (b) and (c) show total SOC energy and MAE as a function of biaxial strain for CrI$_3$ and CrBr$_3$, respectively. The atomic contributions are also plotted. (d) and (e) show the charge density variation (compared to bare atoms) in the strained crystal of CrI$_3$ and CrBr$_3$, respectively. One notes the overlap of the bonding charges under compressive strain, zoomed out for facilitated visualization, and absent for the tensile strain.} 
\end{figure*}

\subsection{Effect of strain}\label{strain}

The responses of monolayer CrI$_3$ and CrBr$_3$ to biaxial strain are further revealing the influence of SOC on the magnetic exchange interaction. Principally, the structural changes upon straining are similar for CrI$_3$ and CrBr$_3$. Cr-Cr distance changes according to the degree of strain applied ($\pm5\%$), however, the Cr-X bond length varies $\sim \pm1\%$. Most of the changes therefore accumulate to the Cr-X-Cr (X = I, Br) angle, varying $\sim \pm5\%$. Such changes in the angle cause distortion in the octahedral coordination, as illustrated in Fig. \ref{f3}(a). Therefore, it is expected that strain induces the interaction of two $spd^2$ bonds of one ligand due to bong-angle modification and leads to modification of the SOC contribution in both Cr and the ligand.

We next calculated the total energy difference and the total SOC energy difference between out-of plane and in-plane spin alignments under biaxial strain. As shown in Figs. \ref{f3} (b) and (c), $E_{\text{SO}}$ of monolayer CrI$_3$ [orange curve in Fig. \ref{f3}(b)] linearly increases under increasing tensile strain, which is similar to its behavior in CrBr$_3$ [green curve in Fig. \ref{f3}(c)]. However, the behavior of the two materials under compressive strain is completely different. CrI$_3$ exhibits parabolic increase with increasing compressive strain, while CrBr$_3$ maintains linear behavior and decreases with increasing compressive strain! The atomic contributions reveal the source of these behaviors. As shown by magenta curve in Fig. \ref{f3}(b), iodine contribution dominates the behavior for both compressive and tensile strain. For CrBr$_3$ in Fig. \ref{f3}(c), in case of the compressive strain, the contributions from Cr and Br are almost equal but have opposite sign, meaning that Br atoms favor to have spin in out-of-plane direction while Cr atoms prefer in-plane spin direction. For tensile strain, Br contribution dominates and both Cr and Br prefer out-of-plane spin alignment. Since the structural changes of both monolayers under strain are similar, one expects similar variation of Cr energies, however, Cr of CrI$_3$ exhibits three times larger energy variation as compared to Cr of CrBr$_3$. This indicates that the anisotropy related with Cr also originates from the bonding electrons rather than Cr-only electrons. In Figs. \ref{f3} (d) and (e) we show the charge density variation for $5\%$ and $-5\%$ strain in both CrI$_3$ and CrBr$_3$. It is clearly seen that in case of compressive $-5\%$ strain the bonding charges of two different bonds of one ligand overlap. This reveals the origin of the interaction under compressive strain. For $5\%$ tensile strain, the bonding charges are clearly separated and exhibit no overlap.  

Our results for behavior of MAE are generally in good agreement with results reported by Ref. \onlinecite{PhysRevB.98.144411}. In case of CrBr$_{3}$ we also obtain that MAE is smallest in case of $-5\%$ strain and it is linearly growing, reaching the maximum for $+5\%$ strain. In case of CrI$_{3}$, however, our results agree with Ref. \onlinecite{PhysRevB.98.144411} only for the compressive strain - with increased compressive strain, anisotropy is growing. On the other hand, with tensile strain, we report completely different behavior of MAE in CrI$_{3}$, compared to Ref. \onlinecite{PhysRevB.98.144411}. There, anisotropy was decreasing with increased tensile strain, while in our study, the behavior is different - anisotropy grows with increased tensile strain.

\begin{figure}[t]
\includegraphics[width=0.65\linewidth]{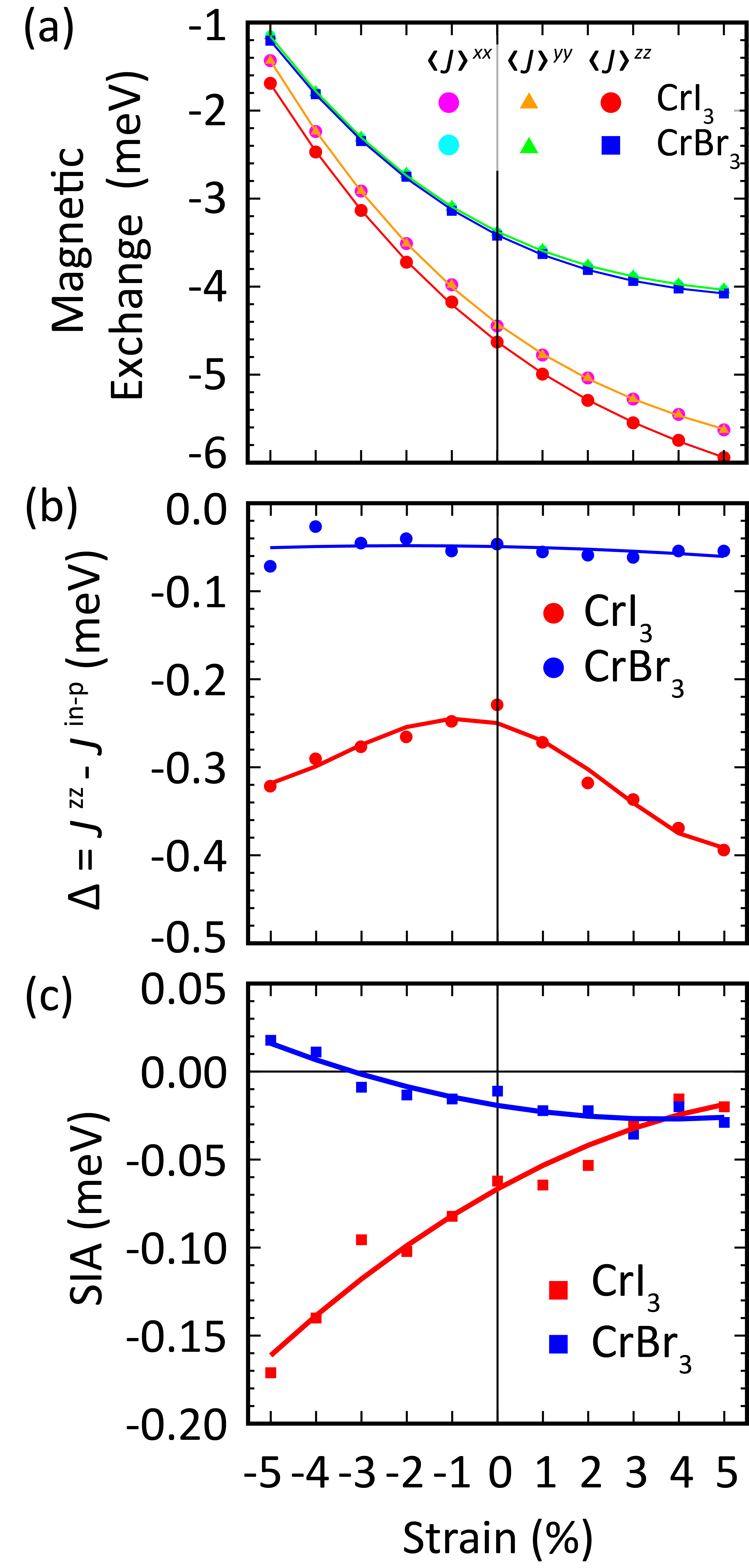}
\caption{\label{f4} The magnetic exchange parameters (a), out-of-plane anisotropy (b), and SIA parameters (c) of monolayer CrI$_3$ and CrBr$_3$ as a function of the biaxial strain.} 
\end{figure}

The variations of exchange parameters and SIA under biaxial strain are also examined based on the considerations related to Table \ref{mag_tab} and Fig. \ref{fx}. The mean values $\langle J \rangle$ of the exchange matrices of three NNs are plotted as a function of strain in Fig. \ref{f4}(a). The ferromagnetic exchange interaction increases with the tensile strain and decreases with compressive strain for both CrI$_3$ and CrBr$_3$, for all components. The variation on the curves of CrI$_3$ is much larger such that the compressive strain almost equalizes CrI$_3$ and CrBr$_3$ in terms of exchange interaction, and the difference between two monolayers increases with tensile strain. These results agree with Ref. \onlinecite{PhysRevB.98.144411} only in case of compressive strain. Namely, both our study and mentioned previous work suggest that with compressive strain, two materials become less FM. However in case of tensile strain, our results suggest that materials are more FM than in pristine case, while in Ref. \onlinecite{PhysRevB.98.144411}, the opposite was suggested. In Fig. \ref{f4}(b), the anisotropy between in-plane and out-of-plane exchange parameters, $\Delta$, is plotted as a function of strain. $\Delta$ of CrI$_3$ increases for increasing either tensile or compressive strain. It is important to note that the behavior of $\Delta$ is consistent with the behavior of MAE in Fig. \ref{f3}, since most of the contribution to MAE comes from the anisotropy of the exchange interaction of the first NN. For CrBr$_3$, the anisotropy slightly increases (decreases) under tensile (compressive) strain, which is also consistent with the behavior of MAE. In Fig. \ref{f4}(c), SIA is presented as a function of strain. SIA in CrI$_3$ exhibits a gradually slowing decrease (in absolute value) when moving from compressive to tensile strain. In CrBr$_3$ on the other hand, SIA exhibits opposite behavior to that of CrI$_3$. It is significant that for compressive strain beyond $-3\%$ SIA parameter becomes positive, indicating preference for in-plane direction of magnetization. Notice that SIA is an order of magnitude smaller than the exchange parameter, therefore its contribution to overall magnetic properties is limited. However, MAE of CrBr$_3$ under high compressive strain is of the same order as SIA, indicating that the drop of MAE is due to the decrease of SIA parameter, while $\Delta$ stays almost constant. 

\subsection{Temperature-dependent magnetization and hysteresis in applied magnetic field}\label{temp-hys}

\begin{figure}[t]
\includegraphics[width=\linewidth]{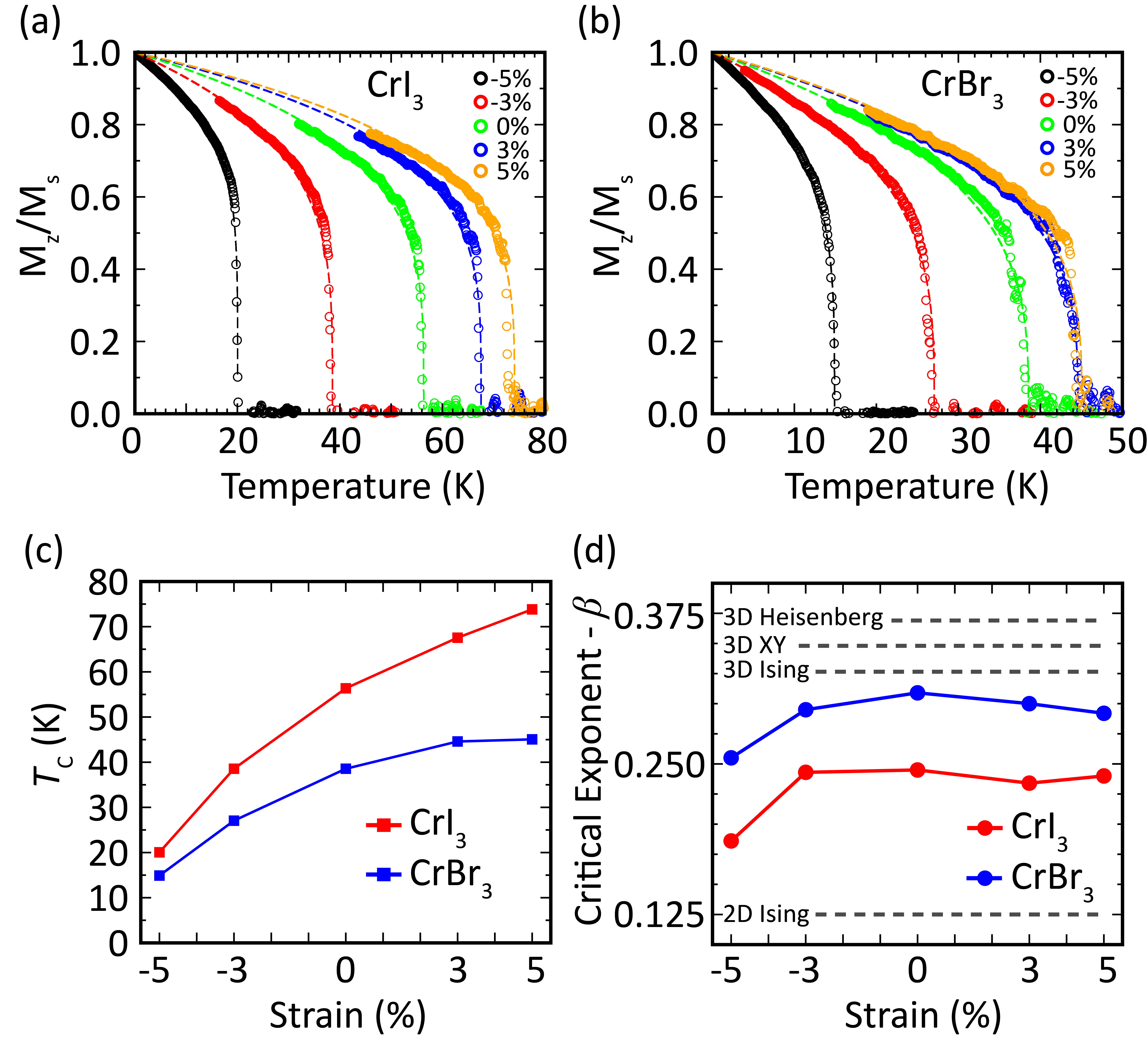}
\caption{\label{f5}  The temperature-dependent magnetization M$_z$/M$_s$($T$) for different amounts of strain applied to monolayer CrI$_3$ (a) and CrBr$_3$ (b). Simulation data are fitted by $(1-T/T_{\text{C}})^\beta$, where $\beta$ is the critical exponent and $T_{\text{C}}$ the Curie temperature. (c) and (d) panels plot the thereby obtained $T_\text{C}$ and $\beta$ as a function of strain, respectively. For comparison, $\beta$ values from different models \cite{gibertini2019magnetic} are shown as dashed lines in (d).} 
\end{figure}

The above-indicated differences in the strength of the magnetic exchange parameters, exchange anisotropy, and SIA parameter between two magnetic monolayers under investigation can be monitored via temperature-dependent magnetization and hysteretic behavior in applied magnetic field, which are both readily experimentally accessible. Having obtained all the parameters to construct the Heisenberg spin Hamiltonian in Eq. \eqref{hamil} for strained monolayers, we next calculate the temperature-dependent magnetization of CrI$_3$ and CrBr$_3$ and the corresponding $T_{\text{C}}$ where the magnetic phase transits from paramagnetic to ferromagnetic state. We used stochastic LLG simulation to obtain the temperature-dependent magnetization M$_z$/M$_s$($T$), where M$_z$ is the out-of-plane magnetization and M$_s$ is the saturation magnetization. As shown in Figs. \ref{f5} (a) and (b), the obtained $T_\text{C}$ of the unstrained monolayers of CrI$_3$ and CrBr$_3$ of 56 K and 38 K is reasonably close to the experimentally obtained values of 45 K and 21 K, respectively. These $T_\text{C}$ values are also consistent with those found in previous works\cite{torelli2018calculating,torelli2019high}, obtained using renormalization spin-wave theory combined with classical MC calculations\footnote{Note that the feeding parameters in those works were obtained by DFT calculations with slightly different parametrization than used in the present work, such as $U=3.5$ eV instead of $U=4$ eV.}. Further, as shown in Fig. \ref{f5}(c), $T_\text{C}$ decreases (increases) under compressive (tensile) strain, mainly due to the previously described strong variation of the exchange parameters with strain. In case of CrBr$_3$, contrary to CrI$_3$, the influence of $\Delta$ and SIA is very small since the variation of those parameters under strain can be considered negligible. One should note that the behavior of $T_\text{C}$ as a function of strain presented here is completely different than that reported in Ref.~\onlinecite{PhysRevB.98.144411} where $T_\text{C}$ is calculated using mean-field theory. Such disagreement is hardly a surprise since in Ref.~\onlinecite{PhysRevB.98.144411} a single exchange parameter is used to determine $T_\text{C}$ in absence of information on anisotropy. Further we note that two monolayer materials have almost equal $T_\text{C}$ at compressive $-5\%$ strain. In the case of tensile strain on the other hand, the difference between $T_{\text{C}}$ of CrI$_3$ and CrBr$_3$ increases with strain. CrBr$_3$ reaches the saturation of $T_{\text{C}}\approx 45$ K at around 3\% strain while CrI$_3$ exhibits $T_{\text{C}}\approx 74$ K at 5\% strain and tends to further increased $T_{\text{C}}$ with further straining.

The behavior of M$_z$/M$_s$($T$) curves is also illustrative of the difference between the two materials. In general, all curves are well fitted by the functional behavior $(1-T/T_{\text{C}})^\beta$, where $\beta$ is the critical exponent. In Fig. \ref{f5}(d), obtained exponents $\beta$ are plotted as a function of strain, together with $\beta$ values from different available models for comparison. In our results, the critical exponent of monolayer CrI$_3$ at all strains can be approximated by $\beta \approx 0.24$. This value suggests that strong out-of-plane anisotropy of CrI$_3$ separates its behavior from those expected by 3D models and sets it closer to the 2D limit. Our value of $\beta$ is also comparable with the value of $\sim0.26$ obtained in Ref. \onlinecite{PhysRevB.97.014420} for bulk CrI$_3$ since the intralayer interactions dominate the magnetic behavior even in bulk CrI$_3$.\footnote{It is interesting to mention that the critical exponent of the 2D XY model is estimated as $\beta \approx 0.23$\cite{bramwell1993magnetization} from the experimental observations on several layered magnets such as BaNi$_2$(PO$_4$)$_2$, Rb$_2$CrCI$_4$ and K$_2$CuF$_4$, which is very close to $\beta$ of CrI$_3$. We believe this to be a coincidence, since 2D XY model is based on the easy-plane anisotropy contrary to the out-of-plane easy-axis of CrI$_3$. It is also known that 2D XY model does not exhibit standard spontaneous phase transition, which should deviate from the behavior we obtained in our simulations at around $T_{\text{C}}$.} CrBr$_3$ with $\beta \approx 0.3$, on the other hand, is much closer to the 3D Ising value ($\beta = 0.3226$) due to weak out-of-plane anisotropy. Our result is validated by (albeit at the lower margin of) $\beta =0.4\pm0.1$ recently obtained from magnetization measurements by Kim \textit{et al.}\cite{kim2019micromagnetometry}.

\begin{figure*}[t]
\includegraphics[width=0.7\linewidth]{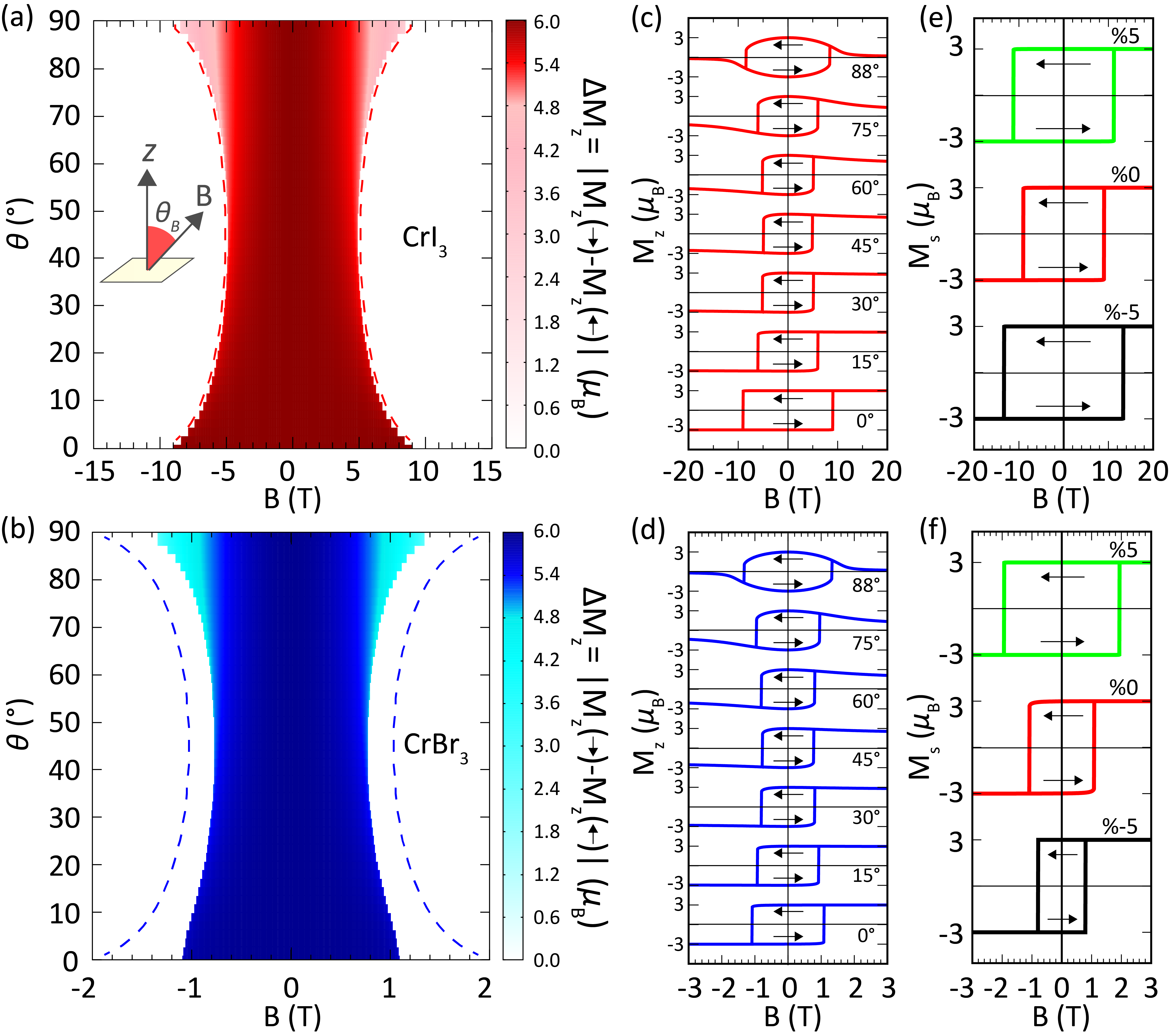}
\caption{\label{f6} The height of the hysteresis loop $\Delta \text{M}_z =|\text{M}_z(\rightarrow)-\text{M}_z(\leftarrow)|$ as a function of the external magnetic field $B$ tilted by angle $\theta_B$ from the out-of-plane direction, for (a) CrI$_3$ and (b) CrBr$_3$. Panels (c) and (d) show the corresponding hysteresis curves obtained for indicated angles $\theta_B$. Panels (e) and (f) highlight the difference in the hysteretic response of two materials when strained (here $\theta_B=0$).} 
\end{figure*}
Finally, motivated by its accessibility by advanced magnetometry,\cite{kim2019micromagnetometry} we calculated the hysteretic behavior of the monolayers under magnetic field $B$ tilted by angle $\theta_B$ from the $z$-axis, for both CrI$_3$ and CrBr$_3$. 

In the simulations, the spin system is initialized at high applied field, where all the spins are aligned to the magnetic field direction. The field is then looped in steps of $0.01$~T and $0.05$~T for CrBr$_3$ and CrI$_3$ respectively, where the magnetization is relaxed by $5\times10^4$ time steps for each value of field. Dipole-dipole interactions have been included in the simulations. In Fig. \ref{f6}(a,b), we show the obtained $\Delta \text{M}_z =|\text{M}_z(\rightarrow)-\text{M}_z(\leftarrow)|$ as a color-map plot where shades of red (CrI$_3$) and blue (CrBr$_3$) color show the deviation of $M_z(B)$ hysteresis loop from the rectangular shape. In Figs. \ref{f6} (c) and (d), we plot the corresponding hysteresis curves for selected tilt angles of the applied magnetic field. One clearly sees that hysteresis loops evolve from sharp rectangular to an oval form as the tilt angle is increased, as was recently shown experimentally by Kim \textit{et al.} for the case of monolayer CrBr$_3$.

In absence of dipolar interactions, the behavior of magnetic spins in external field is captured by minimization of the energy
$E_{TOT}(\theta) = -\rm{MAE}\cos(2\theta) -E_{B}\cos(\theta-\theta_B)$, where $E_{B}$ is the energy associated with the magnetic field, and grows linearly with increasing $B$. The value of the critical, 'switching' field for the given angle $\theta_{B}$, scaled to the switching field for $\theta_{B}=0^\circ$ [$\Gamma=B_{cr}(\theta_B)/B_{cr}(0^\circ)$] is then analytically obtained from the condition that the energy extrema coalesce, leading to equation
\begin{equation}
\frac{(\Gamma^{2}-1)^{3}}{\Gamma^{4}} = -\frac{27}{4} \sin^2(2\theta_{B}).
\end{equation}
This functional dependence of critical field on the tilt angle $\theta_B$ perfectly reproduces the numerically calculated switching field, shown by dashed lines in Figs. \ref{f6}(a,b). Notably, the switching field in absence of dipolar interactions ($\Gamma(\theta_B)$) shows symmetric behavior with respect to $\theta_B=45^\circ$. However, with dipolar interactions included, that symmetry is broken in case of CrBr$_3$, and switching field for $\theta_B=0^\circ$ becomes significantly lower than for $\theta_B\rightarrow90^\circ$. The latter was indeed validated experimentally, in Ref. \onlinecite{kim2019micromagnetometry}. However, dipolar interactions cause no changes in the critical field of CrI$_3$ for any $\theta_B$, which is another important distinction between two materials that could be verified by Hall micromagnetometry. 

One should however note that in our considerations we do not involve finite size effects nor demagnetization, or temperature fluctuations, likely playing an important role in experiment (next to the ever-present defects in the monolayers, that can facilitate magnetic reversal locally). Our simulations explore primarily the effect of the microscopic parameters on the apparent magnetic behavior, with a goal of capturing the intrinsic differences between CrI$_3$ and CrBr$_3$. As a consequence, the switching fields in our simulations are significantly larger than experimentally reported values of $\approx$ 0.15T \cite{huang2017layer} for CrI$_3$ and $\approx$ 0.03T\cite{kim2019micromagnetometry} for CrBr$_3$, for $\theta_B=0$. Having said that, our simulations capture the large ratio between switching fields of the two monolayer materials (approximately 8 and 5 for simulation and experiment, respectively).

To feature yet another experimentally verifiable difference between monolayer CrI$_3$ and CrBr$_3$, we also calculated the hysteresis curves for the strained magnetic monolayers under out-of-plane magnetic field. We recall that MAE and SIA, plotted in Fig. \ref{f3}(b,c) and Fig. \ref{f4}(c), respectively, showed distinctively different behavior in two materials when under strain. As shown in Figs. \ref{f6} (e) and (f), the width of the hysteresis loop strongly changes with the strain. However, while the switching magnetic field of monolayer CrI$_3$ increases under both compressive and tensile strain, the switching field of monolayer CrBr$_3$ is \textit{reduced} by compressive strain while the tensile strain increases the switching field stronger than was the case in CrI$_3$ (measured with respect to the unstrained case). This behavior can thus be mapped on the behavior of exchange anisotropy and SIA under strain, all rooted in the known difference in spin-orbit coupling between the two materials.

\section{Conclusions}\label{conc}
We compared two of the very first ferromagnetic 2D materials, monolayer CrI$_3$ and CrBr$_3$, belonging to the same chromium-trihalide CrX$_3$ family. Although very similar qualitatively in structural and electronic properties (with some quantitative differences such as lattice constant, bond length, and electronic band gap), these materials exhibit strong differences in magnetic properties and their behavior with external stimuli. We attribute these differences to the spin-orbit interaction not only on ligand atoms (X=I, Br) and but also at the bonding orbitals. We show that the energetic preference of the direction of spin of a ligand directly depends on the bonding direction of that particular ligand relative to the spin direction. That means spin-orbit coupling (SOC) energy contribution of an individual ligand can be predicted qualitatively via its coordination and spin direction.

We also present the magnetic exchange parameters for the two monolayers. The mean exchange interaction in CrI$_3$ is larger that one of CrBr$_3$, but the difference between their out-of-plane anisotropy values ($\Delta$) as well as between single-ion anisotropies (SIA) are more than significant. We also clearly demonstrated that the origin of both the out-of-plane anisotropy and the SIA is the spin-orbit interaction, since our analogous analysis without SOC yielded no exchange anisotropy and no SIA. 

By applying biaxial strain, we revealed further differences between monolayer CrI$_3$ and CrBr$_3$. The strain mostly changes the Cr-X-Cr angle instead of the bond length. That results in significant structural distortion of the octahedral units of the monolayers. Magnetic anisotropy energy (MAE) of CrI$_3$ increases under either compressive and tensile strain while MAE of CrBr$_3$ linearly increases (decreases) under increasing tensile (compressive) strain. This difference in the variation of MAE is reflected on the corresponding changes in the out-of-plane anisotropy of the exchange parameters and the SIA parameter. With such differences in obtained parameters for the two materials, we calculated the temperature-dependent magnetization for pristine and the strained monolayers, to reveal much stronger variation with strain of $T_{\text{C}}$ in CrI$_3$ than in CrBr$_3$. The found critical exponent of our $M(T)$ data places CrBr$_3$ virtually in the 3D regime, owing to its low out-of-plane anisotropy, contrary to the strong 2D character of CrI$_3$.

For facilitated direct observation of the reported differences between monolayer chromium-trihalides, and as a direct probe of their magnetic anisotropy, fostered by spin-orbit coupling, we also calculated the behavior of hysteretic magnetization loops as a function of the tilt angle between the applied field and the monolayer plane. We revealed that magnetic behavior of CrBr$_3$ is far more affected by dipolar interactions than is the case in CrI$_3$, but also that the behavior of the switching field with strain is entirely different in two materials, analogously to previously observed differences in MAE and SIA as a function of strain. These findings are yet another proof that even subtle differences in atomic contributions to spin-orbit coupling between two akin materials can lead to rather dissimilar magnetic properties, and can be broadly tuned by gating, straining and heterostructuring of the 2D material. Although sourced in properties at atomistic scale, these differences can clearly manifest in macroscopic observables and are verifiable experimentally (owing to e.g. recent advances in Hall magnetometry). Therefore, tailored solutions for spatially engineered spin-orbit coupling in magnetic monolayers present an attractive roadmap towards advanced spintronic and magnonic nanocircuitry.

\begin{acknowledgments}
This work was supported by the Research Foundation-Flanders (FWO-Vlaanderen) and the Special Research Funds of the University of Antwerp (TOPBOF). The computational resources and services used in this work were provided by the VSC (Flemish Supercomputer Center), funded by Research Foundation-Flanders (FWO) and the Flemish Government -- department EWI.
\end{acknowledgments}

\end{document}